# Various magnetism of the compressed antiferromagnetic topological insulator EuSn$_2$As$_2$

Hualei Sun[1,*], Cuiqun Chen[1], Yusheng Hou[1], Yu Gong[2], Mengwu Huo[1], Lisi Li[1], Jia Yu[1], Wanping Cai[1], Naitian Liu[1], Ruqian Wu[3], Dao-Xin Yao[1], and Meng Wang[1,†]

[1]*School of physics, Sun Yat-Sen University, Guangzhou, Guangdong 510275, China*

[2]*Beijing Synchrotron Radiation Facility, Institute of High Energy Physics, Chinese Academy of Sciences, Beijng 100049, China*

[3]*Department of Physics and Astronomy, University of California, Irvine, California 92697-4575, USA*

*\* sunhlei@mail.sysu.edu.cn*

*† wangmeng5@mail.sysu.edu.cn*

We report a comprehensive high-pressure study on the antiferromagnetic topological insulator EuSn$_2$As$_2$ up to 21.1 GPa through measurements of synchrotron x-ray diffraction, electrical resistance, magnetic resistance, and Hall transports combined with first-principles calculations. No evident trace of a structural phase transition is detected. The Néel temperatures determined from resistance are increased from 24±1 to 77±8 K under pressure, which is resulted from the enhanced magnetic exchange couplings between Eu$^{2+}$ ions yielded by our first-principles calculations. The negative magnetoresistance of EuSn$_2$As$_2$ persists to higher temperatures accordingly. However, the enhancement of the observed Néel temperatures deviates from the calculations obviously above 10.0 GPa. In addition, the magnitude of the magnetoresistance, the Hall coefficients, and the charge carrier densities show abrupt changes between 6.9 to 10.0 GPa. The abrupt changes probably originate from a pressure induced valence change of Eu ions from a divalent state to a divalent and trivalent mixed state. Our results provide insights into variation of the magnetism of EuSn$_2$As$_2$ and similar antiferromagnetic topological insulators under pressure.

## I. INTRODUCTION

Magnetic topological insulators (TIs) have attracted wide attention for their fascinating topological quantum states such as the quantum anomalous Hall effect, axion insulator states, chiral Majorana fermions, and the quantized topological magneto-electric effect.[1–6] These novel features fertilize magnetic TIs potential promising applications such as spintronics and topological quantum computation.[7,8] Magnetic orders and the nontrivial electronic topology of the TIs interact intimately. Effective regulation of the magnetism and elevating the magnetic transition temperature for the magnetic TIs are hence crucial for achieving exotic topological quantum phenomena.[9,10]

EuSn$_2$As$_2$ crystallizes in the trigonal *R*-3*m* space group and has intralayer magnetic Eu$^{2+}$ ions sandwiched by two honeycomb [SnAs]$^{2-}$ layers as shown in the inset sketch in Fig. 1b. The neighboring honeycomb layers interact via van der Waals (vdW) force.[11] At ambient pressure, EuSn$_2$As$_2$ undergoes a paramagnetic (PM) phase to an antiferromagnetic (AFM) phase transition at $T_N \approx 24$ K and exhibits positive magnetoresistance at 60K.[12] Below $T_N$, the intralayer magnetic exchange coupling between Eu$^{2+}$ ions is ferromagnetic (FM), while the interlayer coupling is AFM, leading to an *A*-type AFM order. It is recently proved that EuSn$_2$As$_2$ is an AFM TI with no observed gap in the Dirac surface state at low temperatures by both angel-resolved photoemission spectroscopy (ARPES) and density functional theory (DTF) calculations.[13] In the PM phase, EuSn$_2$As$_2$ has both inversion and time-reversal symmetries. When EuSn$_2$As$_2$ transforms into the AFM phase, the time-reversal symmetry is broken, while the inversion symmetry is preserved. Through band calculations, it is shown that EuSn$_2$As$_2$ is a strong TI in the PM phase with the topological invariant $Z_2=1$, and an axion insulator in the AFM phase with the topological invariant $Z_4=2$.[14]

Pressure is a pure and effective method to tune lattice parameters that are crucial for magnetic structure, magnetic exchange coupling, electronic topological state, and even valence state. For the well-known intrinsic TI $MnBi_2Te_4$, the Néel temperature is first raised to the maximum (~29.6 K) at around 2 GPa and then decreased until vanishing completely at around 7 GPa.[15] It has also been suggested that pressure can elevate the magnetic transition temperature in the AFM insulator candidate $EuIn_2As_2$ significantly through strengthening the magnetic exchange couplings.[13] Note that Eu ions have two different valences, the divalent $Eu^{2+}$ and the trivalent $Eu^{3+}$. $Eu^{2+}$ has the electronic state of $4f^7$ and a large magnetic moment of $7\mu_B$ (S=7/2), while $Eu^{3+}$ has the electronic state of $4f^6$ and is nonmagnetic. Since the nonmagnetic $Eu^{3+}$ has a smaller volume than that of the magnetic $Eu^{2+}$, pressure is expected to induce a valence state transformation from $Eu^{2+}$ to $Eu^{3+}$. Actually, such transformation has been suggested in previous high-pressure electrical resistivity, Mossbauer spectroscopy, magnetic susceptibility, x-ray absorption spectroscopy and x-ray magnetic circular dichroism studies in several Eu-based materials.[16–20] The specific features of Eu ions in AFM TIs against pressure thus call for comprehensive studies such as in $EuSn_2As_2$ because the magnetic properties have an important role in determining its electronic topology.

In this work, we combined a systematic experimental high-pressure study with first-principles calculations on the magnetic TI $EuSn_2As_2$. We find that $EuSn_2As_2$ retains the rhombohedral symmetry up to 21.1 GPa and no structural phase transition is observed. From 0.3 to 18.0 GPa, the $T_N$ increases from 24±1 to 77±8 K and the temperature range of the negative magnetoresistance extends towards to higher temperatures. According to our first-principles calculations, the increase of the magnetic transition temperature originates from enhancement of the intralayer FM exchange couplings between $Eu^{2+}$ ions under pressure. The effect of magnetoresistance becomes quite weak above 10.0 GPa. The Hall coefficient ($R_H$) decreases monotonically as the pressure increases from zero to 10 Gpa, but it remains nearly a constant when the pressure goes above 10 Gpa. The charge carrier density (n) increases significantly between 6.9 and 10.0 GPa accordingly. The evolutions of $R_H$ and n, and the deviation of $T_N$ from the theoretical values probably result from the pressure induced valence transition of Eu ions from a divalent state of $Eu^{2+}$ to a trivalent state of $Eu^{3+}$.

## II. EXPERIMENTAL AND DFT COMPUTATIONAL DETAILS

Bulk single crystals of $EuSn_2As_2$ were grown by the self-flux method. High purity starting metals were mixed and grown as previous reports.[12,21] The structure of $EuSn_2As_2$ was confirmed by single-crystal x-ray diffraction (XRD). Magnetic and electrical measurements were taken on a physical property measurement system (PPMS, Quantum Design). High-pressure synchrotron radiation XRD patterns of $EuSn_2As_2$ were collected at room temperature with an x-ray wavelength of 0.6199 Å. A symmetric diamond anvil cell (DAC) with a pair of 400-μm-diameter culets was used. A sample chamber with a diameter of 150 μm was drilled by laser in a pre-indented steel gasket. $EuSn_2As_2$ single crystals were ground into fine powers and compressed into a pellet with 100 μm in diameter and 20 μm in thickness. The pellet was loaded in the middle of the sample chamber and silicone oil was used as a pressure-transmitting medium. A ruby sphere was also loaded into the sample chamber and pressure was determined by measuring the shift of the fluorescence wavelength of the ruby sphere.

High-pressure electrical transport measurements on $EuSn_2As_2$ single crystals were carried out using a miniature DAC made from Be-Cu alloy on a PPMS. Diamond anvils with a 400-μm culet were used, and the corresponding sample chamber with a diameter of 150 μm was made in the insulating gasket achieved by cubic boron nitride and epoxy mixture. NaCl powders were employed as the pressure-transmitting medium, providing a quasi-hydrostatic environment. Pressure was also calibrated using the ruby fluorescence shift at room temperature. The standard four-probe technique was adopted in these measurements.

Our first-principles calculations were carried out based on density function theory. The Vienna Ab *initio* Simulation Package (VASP) with the generalized gradient approximation was employed.[22,23] We adopted the projector-augmented wave pseudopotentials to describe the core-valence interaction and an energy cutoff of 400 eV for the basis expansion.[24,25] The experimentally measured lattice constants for different pressures were used in our structural relaxations and only the inner atomic positions were relaxed until the force acting on each atom was smaller than 0.01 eV/Å. The interlayer vdW interaction is described by the nonlocal vdW functional of optB86b-vdW.[26,27] To account the strong correlation among the $f$ electrons of $Eu^{2+}$ ions, we used $U$=6.5 eV and $J$=1.0 eV, respectively. The Néel temperature of the pressurized $EuSn_2As_2$ was obtained through parallel tempering Monte Carlo (MC) simulations.[28,29]

### III. RESULTS
### A. High-pressure structure of $EuSn_2As_2$

Figure 1 displays investigations of the crystal structure under pressure up to 21.1 GPa at room temperature. All the diffraction peaks of the high-pressure XRD can be indexed by the trigonal *R*-3*m* space group. No structural transition is observed within the pressure range we measured. The enhancement of the diffraction peak at $2\theta$=14.5° that corresponds to (1 0 7) above 13.0 GPa could be attributed to the reorientation of the powder crystals under pressure. The experimental observed $2\theta$ position of the diffraction peak indexed by (1 0 7) is consistent with the simulation at every individual pressure. Figure 1b shows pressure dependences of the lattice constants and cell volume derived from Fig. 1a. The lattice constants and cell volume decrease smoothly under pressure, revealing preservation of the original crystal structure in the measured pressure range. The *c* axis which is perpendicular to the vdW layers is found to be more compressible. This is similar with $EuIn_2As_2$ below 17 GPa. Above 17 GPa, $EuIn_2As_2$ undergoes a crystalline-to-amorphous phase transition.[13]

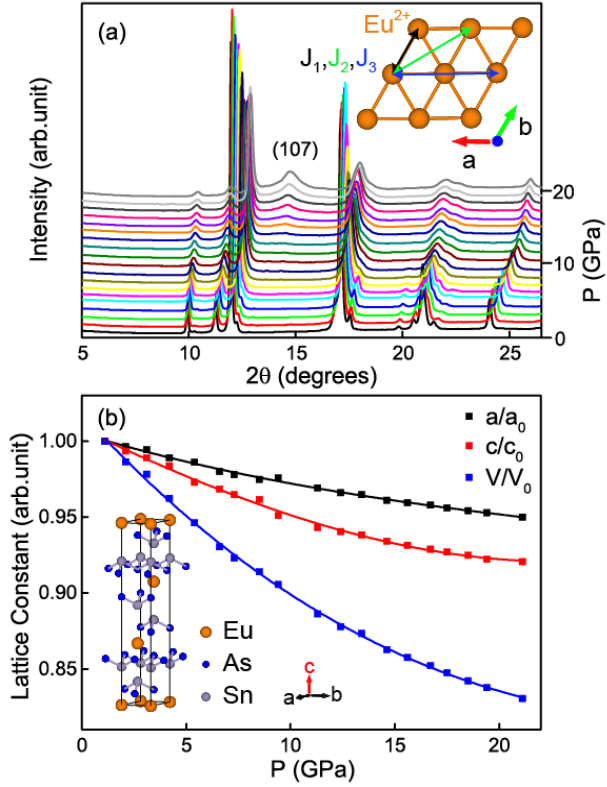

**Figure 1** (a) High pressure XRD patterns of EuSn$_2$As$_2$ from 1.1 to 21.1 GPa. The diffraction data has been scaled to the background (shown on the left $y$ axis) and the offsets are set to the measured pressures (shown on the right $y$ axis). (b) Pressure dependence of the lattice constants and cell volumes normalized as the ambient pressure values. The data were obtained from high-pressure XRD. The insets in (a) and (b) show the intralayer exchange paths of Eu$^{2+}$ and the crystal structure of EuSn$_2$As$_2$, respectively.

## B. High-pressure resistivity of EuSn$_2$As$_2$

In order to investigate the electrical properties of EuSn$_2$As$_2$ under pressure, we performed high-pressure resistivity measurements. Figures 2a and 2b show temperature dependence of the resistivities at various pressures on the absolute and normalized units, respectively. The resistivities above 80 K were fitted to $\rho = \rho_0 + AT^n$ in Fig. 2a, where $\rho_0$ is the residual resistivity, $A$ is the temperature coefficient, and $n$ is the exponent. The values of $n$ vary between 0.92±0.01 (0.3 GPa) and 1.67±0.01 (18.0 GPa), suggesting a non-Fermi-liquid behavior. At 0.3 and 4.0 GPa, the resistivities bend upwards at low temperature, which is similar to that of measured at ambient pressure and has been attributed to the Kondo scattering of spin disorders.[31] As increasing pressure, the upwarping on the resistivity at low temperature disappears, indicating restraining of the Kondo effect.

The Néel temperatures of the AFM transition under pressure could be identified from the normalized $\rho(T)/\rho(150K)$ and derivatives of the resistivities in the inset figure in Fig. 2b. The $T_N$ at 0.3 GPa is close to 24 K that of at ambient pressure. As the pressure increases, the derived $T_N$s increase from 24±1 at 0.3 GPa to 77±8 K at 18.0 GPa, as shown in Fig. 2c. As a comparison, we present the evolution of the Néel temperatures of EuIn$_2$As$_2$ in Fig. 2c. The $T_N$ of EuSn$_2$As$_2$ increases faster than that of EuIn$_2$As$_2$ below 10.0 GPa and statures for the higher pressures.

To gain insight into the enhanced $T_N$ in compressed EuSn$_2$As$_2$, we calculated its exchange couplings based on the following Heisenberg model:

$$H = J_1 \sum_{\langle ij \rangle} \mathbf{S}_i \cdot \mathbf{S}_j + J_2 \sum_{\langle\langle ij \rangle\rangle} \mathbf{S}_i \cdot \mathbf{S}_j + J_3 \sum_{\langle\langle\langle ij \rangle\rangle\rangle} \mathbf{S}_i \cdot \mathbf{S}_j + J_z \sum_{\langle ij \rangle \in \text{interlayer}} \mathbf{S}_i \cdot \mathbf{S}_j \quad (1).$$

In Eq. (1), $J_1$, $J_2$, and $J_3$ are the first-, second- and third-nearest neighbor intralayer exchange couplings, respectively, as shown in the inset of Fig. 1a, and the last term describes the interlayer exchange coupling. First of all, our calculations show that the AFM order has a lower energy than the FM order when the pressure ranges from 0 to 18.0 Gpa, consistent with the experimental observations. From Table 1, one can see that the dominating exchange coupling is the FM $J_1$ and $J_1$ increases as the pressure increasing. This is understandable because the distance between the intralayer $Eu^{2+}$ ions decreases under pressure. Considering that the interlayer AFM $J_z$ is much weaker than the intralayer exchange coupling, we do not take $J_z$ into account in the MC simulations for simplicity and only use the critical temperature of one monolayer of $Eu^{2+}$ ions to estimate the $T_N$ of the compressed $EuSn_2As_2$. As shown in Fig. 2c, the calculated $T_N$s are consistent with the experimental observation, especially for pressures below 10 GPa. However, a clear discrepancy between the measured and calculated $T_N$s appears for higher pressures. This discrepancy could be attributed to the valence transition of Eu ions from $Eu^{2+}$ to $Eu^{3+}$, because the $Eu^{3+}$ ions are nonmagnetic and have a smaller volume. The conduction electrons that have not been considered in the Heisenberg model would also modify the calculated $T_N$s. Overall, the enhancement of $T_N$s mainly results from the strengthening of the intralayer exchange couplings.

**Table 1** Magnetic exchange couplings $J_1$, $J_2$, $J_3$, and $J_z$ (in units of meV) for the compressed $EuSn_2As_2$. The energy difference $\Delta E$ is defined as $\Delta E = E_{AFM} - E_{FM}$ where $E_{AFM}$ and $E_{FM}$ are the energies of the compressed $EuSn_2As_2$ with the AFM and FM orders, respectively. Here negative (positive) exchange coupling parameters correspond to the FM (AFM) Heisenberg exchange interactions.

| Pressure (Gpa) | $J_1$ | $J_2$ | $J_3$ | $J_z$ | $\Delta E$ (meV/f.u.) |
|---|---|---|---|---|---|
| 0.0 | -1.98 | 0.20 | -0.16 | 0.19 | -1.15 |
| 4.2 | -2.92 | 0.21 | -0.32 | 0.17 | -1.00 |
| 8.5 | -3.98 | -0.05 | -0.55 | 0.21 | -1.24 |
| 12.4 | -5.08 | -0.38 | -0.92 | 0.19 | -1.14 |
| 18.0 | -6.35 | -0.63 | -1.19 | 0.24 | -1.45 |

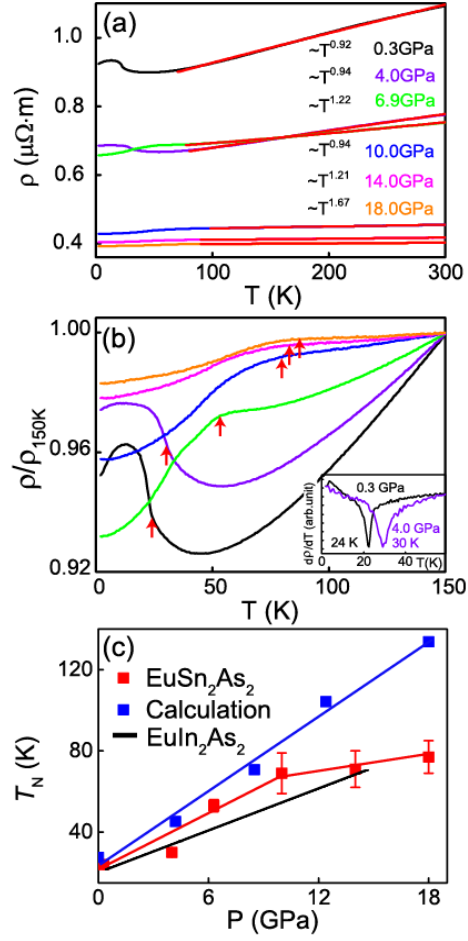

**Figure 2** (a) Temperature dependence of the resistivity at pressures from 0.3 to 18.0 GPa. The red lines present the fitted curves using $\rho = \rho_0 + AT^n$. (b) Temperature dependence of the normalized $\rho(T)/\rho(150K)$ under various pressures from 0.3 to 18.0 GPa. The arrows mark the Néel temperatures at different pressures. The inset in (b) shows the first derivative of the resistivity with respect to temperature and the singular points correspond to the Néel temperatures. (c) The derived Néel temperatures at different pressures. The red and blue lines are linear fittings of the experimentally measured and first-principles calculated Néel temperatures for EuSn$_2$As$_2$, respectively. For comparison, the Néel temperatures of EuIn$_2$As$_2$ under pressure are shown as the black line.

### C. High-pressure magnetoresistance of EuSn$_2$As$_2$

To get better understanding on the magnetic properties of EuSn$_2$As$_2$ as a function of pressure, we measured magnetoresistance under various pressures from 3 to 150 K, as presented in Figs. 3 and 4. In this temperature range, EuSn$_2$As$_2$ goes through the AFM state to a PM state. We define *MR* as

$$MR = \frac{\rho(H) - \rho(0)}{\rho(0)} \times 100\% \quad (2).$$

where ρ(H) and ρ(0) are the resistivities measured at magnetic field μ$_0$H and zero field, respectively. During the measurements, the current is in the *ab* plane, and μ$_0$H is along the *c* direction. The magnetic and electric fields maintain perpendicularly to each other. Hence, the observed negative *MR* is resulted from the spin scattering of the magnetic Eu$^{2+}$ ions instead of the chiral anomaly.[32]

Figures 3a-3f show the *MR* from 3 to 60 K at pressures in the range of 0.3 to 18.0 GPa. The overall *MR* exhibits a significant reduction in the high-pressure range (10.0 to 18.0 GPa) compared with that of the low-pressure range (0.3 to 10.0 GPa). To quantitatively track the evolution of the magnetoresistance, we integrate the *MR* over the magnetic field from -9 to 9T and present them in Fig. 3g. The integrals of the *MR* at 3, 10, 20 and 30 K increase from 0.3 to 4.0 GPa and decrease quickly from 4.0 to 10.0 GPa, then diminish gradually from 10.0 to 18.0 GPa. For the integrals of the *MR* at 40, 50 and 60 K, they reach maxima at 6.9 GPa, then decline gradually with further increasing the pressure. In Figs. 3a-3c, the *MR* has sharp kinks against magnetic field at 3 and 10 K, which could be attributed to spin-flip transitions. At 3 K, the magnetic fields of the spin-flip transition are 3.32 T at 0.3 GPa, 1.75 T at 4.0 GPa, and 0.95 T at 6.9 GPa, respectively.

The *MR* at 80, 100, and 150 K for various pressures is plotted in Fig. 4. Both the AFM order and AFM correlation could induce negative magnetoresistance. As shown in Fig. 4a, the temperatures are well above $T_N$ = 24±1 K at 0.3 GPa, and the *MR* is positive.

For the pressure of 4.0 GPa with $T_N$ = 30±2 K, the negative *MR* at 80 K in Fig. 4b may result from enhancement of the AFM correlation under pressure. With further increasing the pressure, the appearance of the AFM correlation extends to higher temperatures. The increase of the AFM correlation temperature range consists with the negative *MR* at 80, 100, 150 K and 10.0, 14.0, and 18.0 GPa as shown in Figs. 4d-4f. However, the progressive decrease of the magnitude of the *MR* at high pressures is consistent with the transformation of the magnetic $Eu^{2+}$ to nonmagnetic $Eu^{3+}$.

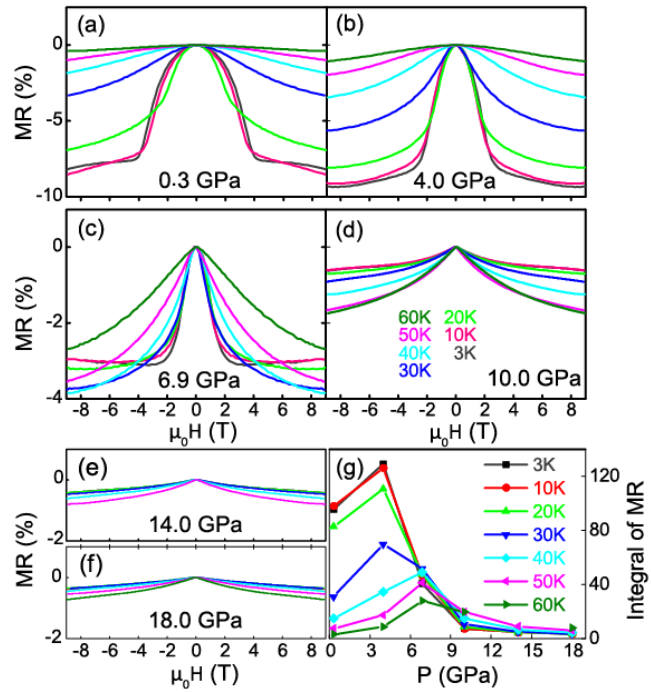

**Figure 3** (a)-(f) *MR* of $EuSn_2As_2$ single crystals measured from 3 to 60 K under various pressures. The *y* range of each two adjacent pressures is set to be the same. (g) Integrals of *MR* over the magnetic field from -9 to 9 T as a function of pressure from 3 to 60 K.

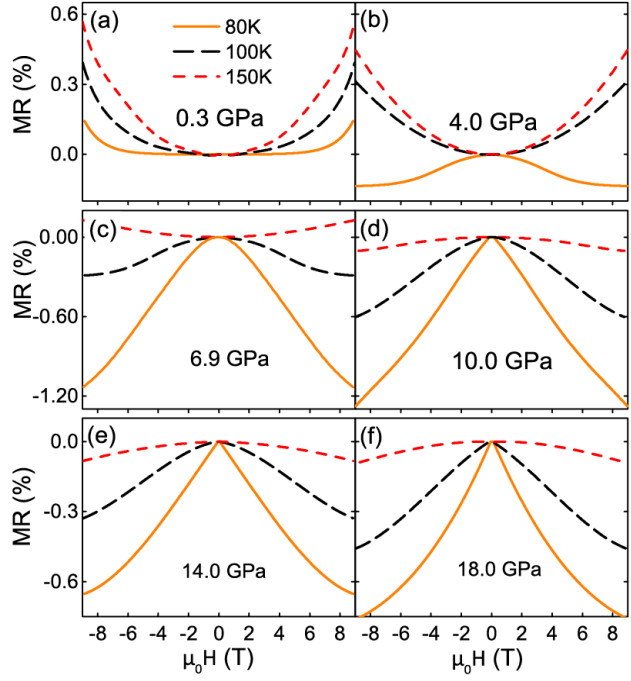

**Figure 4** MR of EuSn$_2$As$_2$ single crystals measured at 80, 100, and 150 K under pressures of (a) 0.3, (b) 4.0, (c) 6.9, (d) 10.0, (e) 14.0, and (f) 18.0 GPa.

### D. High-pressure Hall resistance of EuSn$_2$As$_2$

In order to investigate the evolution of the electronic properties under pressure, we measured the Hall resistance $R_{xy}$ at pressures ranging from 0.3 to 18.0 GPa at 30 and 60 K. The results are displayed in Fig. 5a. At high magnetic field, $(R_{xy}^+ - R_{xy}^-)/2$ varies approximately linearly against the magnetic field. The low magnetic field Hall resistance at 4.0 and 6.9 GPa is non-linear in relation with the magnetic field at 30 K. This is probably due to the asymmetry of the electrodes and the fact that magnetoresistance effect is more pronounced at 4.0 and 6.9 GPa, as shown in Fig. 3g. The Hall coefficient and carrier density are calculated from the standard method using the slope of $(R_{xy}^+ - R_{xy}^-)/2$ in the high magnetic field, where $(R_{xy}^+ - R_{xy}^-)/2$ is approximately linear against the magnetic field. The Hall coefficient remains positive, revealing that the majority of the carriers are holes. Figure 5b displays the Hall coefficient and the density of the carriers, which show the same trend at 30 and 60 K. The Hall coefficient decreases monotonically up to 10.0 GPa then holds a constant, resulting in a transition-like change between 6.9 and 10.0 GPa for the density of carriers. The density of carriers above 10.0 GPa accumulates around 6.6 (30 K) and 10.4 times (60 K) of that at 0.3 GPa.

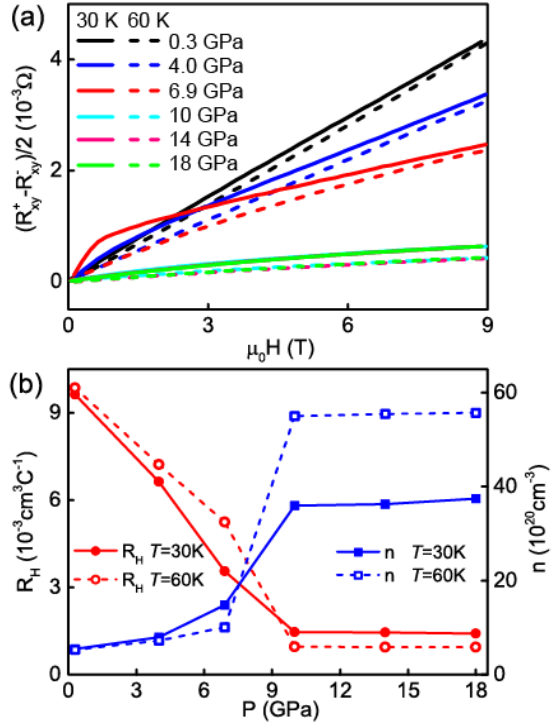

**Figure 5** (a) Hall resistance $(R_{xy}^+ - R_{xy}^-)/2$ at 30 and 60 K under various pressures. $R_{xy}^+$ and $R_{xy}^-$ were measured with a positive and negative magnetic field, respectively. The direction of the magnetic field is along the *c* axis. (b) Pressure dependence of Hall coefficients (black) and carrier densities (blue) measured at 30 (solid dots) and 60 K (hollow dots). The Hall coefficients and carrier densities are calculated within the high magnetic fields range, where $(R_{xy}^+ - R_{xy}^-)/2$ is approximately linear in relation with the magnetic field.

## IV. DISCUSSION AND SUMMARY

The high pressure XRD measurements reveal that EuSn$_2$As$_2$ preserves the rhombohedral structure up to 21.1 GPa. However, there are obvious changes between 6.9 and 10.0 GPa on the Néel temperature, magnetoresistance, Hall coefficient, and carrier density. Especially, when the pressure is above 10.0 GPa, the experimentally measured Néel temperatures $T_N$s distinctly deviate from the calculations. The negative magnetoresistance is suppressed quickly between 4.0 and 10.0 GPa, and becomes weak above 10.0 GPa. The carrier density is enhanced significantly above 6.9 GPa, and saturates above 10.0 GPa. As revealed by our calculations, the enhancement of the Néel temperature under pressure up to 18.0 GPa is mainly due to the increase of the intralayer magnetic exchange couplings of Eu$^{2+}$, similar with the observation of a pressure induced enhancement of the $T_N$ in EuSnP and EuIn$_2$As$_2$.[13,33] While the deviation of the observed $T_N$ from the calculations and the suppressed magnetoresistance above 10.0 GPa suggest involvement of other mechanisms. We note pressure prefers to induce the valence state of Eu ions changing from Eu$^{2+}$ to Eu$^{3+}$ since the nonmagnetic Eu$^{3+}$ ion has a smaller volume. The pressure-induced transformation from Eu$^{2+}$ to Eu$^{3+}$ has been observed in EuFe$_2$As$_2$, EuNi$_2$Ge$_2$, EuPd$_2$Ge$_2$, EuPt$_2$Si$_2$, and EuRh$_2$Si$_2$.[17–20,34] Taking EuFe$_2$As$_2$ as an example, the Eu $L_3$-edge x-ray absorption spectrum exhibits a single-peak structure that corresponds to Eu$^{2+}$ ions at low pressure. Above ~9 GPa, the spectrum consists of two resolved peaks, indicating a mixed-valent state of both Eu$^{2+}$ and Eu$^{3+}$. The transformation from Eu$^{2+}$ to nonmagnetic Eu$^{3+}$ is consistent with our observations in EuSn$_2$As$_2$. The $T_N$ deviates from the calculations and the magnetoresistance is suppressed as the fraction of the nonmagnetic Eu$^{3+}$ increases. The change of the carrier density is a cooperative effect of the chemical potential and the transformation of $Eu^{2+}$ and $Eu^{3+}$ under pressure.

During the preparation of this work, we noticed reports of the observation of superconductivity in $EuSn_2As_2$ with a superconducting (SC) transition temperature $T_c$ = 4.8 K at ambient pressure[35] and $T_c$ = 4 K at 15 GPa[36], respectively. We observed superconductivity in a sample of $EuSn_2As_2$ with a small amount of Sn impurity. However, no superconductivity is observed in the sample we present in this work for pressures up to 18.0 GPa. Further research is still needed to identify the origin of the superconductivity in $EuSn_2As_2$ samples.

In summary, we have studied the structure and electronic transport properties under pressure. We find that the temperatures of the AFM transition and negative magnetoresistance are obviously enhanced under pressure, and an abrupt change of the Hall coefficient and carrier density at ~10.0 GPa. No structural transition is observed up to 21.1 GPa. However, the experimentally measured Néel temperature $T_N$ deviates from the linearly increased $T_N$ as revealed by our calculations. Besides, the magnitude of the magnetoresistance is suppressed, conflicting with the enhanced intralayer magnetic exchange couplings under pressure. We argue that they are a cooperative effect of the enhanced magnetic exchange couplings and a transformation of the Eu irons from $Eu^{2+}$ to $Eu^{3+}$ under pressure. Our work provides comprehensive insights into the varied magnetism of the antiferromagnetic topological insulator $EuSn_2As_2$ under pressure, also benefit understanding on the studies of similar antiferromagnetic topological insulators.

## V. Acknowledgements

Work at Sun Yat-Sen University was supported by the National Natural Science Foundation of China (Grants No. 11904414, 11904416, 11574404, 11974432), National Key Research and Development Program of China (Grant No. 2019YFA0705702, 2018YFA0306001, 2017YFA0206203), GBABRF-2019A1515011337，the Fundamental Research Funds for the Central Universities (Grant No. 18lgpy73).

1. Otrokov, M. M. *et al.* Unique Thickness-Dependent Properties of the van der Waals Interlayer Antiferromagnet $MnBi_2Te_4$ Films. *Phys. Rev. Lett.* **122**, 107202 (2019).

2. Gong, Y. *et al.* Experimental Realization of an Intrinsic Magnetic Topological Insulator. *Chinese Phys. Lett.* **36**, 076801 (2019).

3. Liu, Q., Liu, C. X., Xu, C., Qi, X. L. & Zhang, S. C. Magnetic impurities on the surface of a topological insulator. *Phys. Rev. Lett.* **102**, 156603 (2009).

4. Li, J. *et al.* Intrinsic magnetic topological insulators in van der Waals layered $MnBi_2Te_4$-family materials. *Sci. Adv.* **5,** eaaw5685 (2019).

5. Wu, J. *et al.* Natural van der Waals heterostructural single crystals with both magnetic and topological properties. *Sci. Adv.* **5**, eaax9989 (2019).

6. Tokura, Y., Yasuda, K. & Tsukazaki, A. Magnetic topological insulators. *Nat. Rev. Phys.* **1**, 126–143 (2019).

7. Šmejkal, L., Mokrousov, Y., Yan, B. & MacDonald, A. H. Topological antiferromagnetic spintronics. *Nat. Phys.* **14**, 242–251 (2018).

8. Mogi, M. *et al.* A magnetic heterostructure of topological insulators as a candidate for an axion insulator. *Nat. Mater.* **16**, 516–521 (2017).

9. He, Q. L. *et al.* Tailoring exchange couplings in magnetic topological-insulator/antiferromagnet heterostructures. *Nat. Mater.* **16**, 94–100 (2017).

10. Chang, C. Z. *et al.* Experimental observation of the quantum anomalous Hall effect in a magnetic topological Insulator. *Science.* **340**, 167–170 (2013).

11. Arguilla, M. Q. *et al.* EuSn2As2: An exfoliatable magnetic layered Zintl-Klemm phase. *Inorg. Chem. Front.* **4**, 378–

386 (2017).

12. Chen, H. C. *et al.* Negative Magnetoresistance in Antiferromagnetic Topological Insulator $EuSn_2As_2$. *Chinese Phys. Lett.* **37**, 047201 (2020).

13. Yu, F. H. *et al.* Elevating the magnetic exchange coupling in the compressed antiferromagnetic axion insulator candidate $EuIn_2As_2$. *Phys. Rev. B* **102**, 180404(R) (2020).

14. Li, H. *et al.* Dirac Surface States in Intrinsic Magnetic Topological Insulators $EuSn_2As_2$ and $MnBi_{2n}Te_{3n+1}$. *Phys. Rev. X* **91**, 041039 (2019).

15. Chen, K. Y. *et al.* Suppression of the antiferromagnetic metallic state in the pressurized $MnBi_2Te_4$ single crystal. *Phys. Rev. Mater.* **3**, 094201 (2019).

16. Debessai, M., Matsuoka, T., Hamlin, J. J., Schilling, J. S. & Shimizu, K. Pressure-induced superconducting state of europium metal at low temperatures. *Phys. Rev. Lett.* **102**, 197002 (2009).

17. Matsubayashi, K. *et al.* Pressure-induced changes in the magnetic and valence state of $EuFe_2As_2$. *Phys. Rev. B.* **84**, 024502 (2011).

18. Mitsuda, A., Hamano, S., Araoka, N., Yayama, H. & Wada, H. Pressure-induced valence transition in antiferromagnet $EuRh_2Si_2$. *J. Phys. Soc. Japan* **81**, 023709 (2012).

19. Hesse, H. J. & Wortmann, G. $^{151}$Eu-Mössbauer study of pressure-induced valence transitions in $EuM_2Ge_2$ (M=Ni, Pd, Pt). *Hyperfine Interactions* **93**, 1499–1504 (1994).

20. Ikeda, Y. *et al.* Transport properties under high pressure of antiferromagnet $EuPt_2Si_2$ with unstable Eu valence. *J. Magn. Magn. Mater.* **310**, 62–64 (2007).

21. Goforth, A. M., Klavins, P., Fettinger, J. C. & Kauzlarich, S. M. Magnetic properties and negative colossal magnetoresistance of the rare earth Zintl phase $EuIn_2As_2$. *Inorg. Chem.* **47**, 11048–11056 (2008).

22. Perdew, J. P., Burke, K. & Ernzerhof, M. Generalized gradient approximation made simple. *Phys. Rev. Lett.* **77**, 3865 (1996).

23. Vargas-Hernández, R. A. Bayesian Optimization for Calibrating and Selecting Hybrid-Density Functional Models. *J. Phys. Chem. A* **124**, 4053–4061 (2020).

24. Blöchl, P. E. Projector augmented-wave method. *Phys. Rev. B* **50**, 17953 (1994).

25. Joubert, D. From ultrasoft pseudopotentials to the projector augmented-wave method. *Phys. Rev. B* **59**, 1758 (1999).

26. Klime, J., Bowler, D. R. & Michaelides, A. Van der Waals density functionals applied to solids. *Phys. Rev. B* **83**, 195131 (2011).

27. Klimeš, J., Bowler, D. R. & Michaelides, A. Chemical accuracy for the van der Waals density functional. *J. Phys. Condens. Matter* **22**, 022201 (2010).

28. Lou, F. *et al.* PASP: Property analysis and simulation package for materials. *J. Chem. Phys.* **154**, 114103 (2021).

29. Hukushima, K. & Nemoto, K. Exchange Monte Carlo Method and Application to Spin Glass Simulations. *J. Phys. Soc. Jpn* **65**, 1604–1608 (1996).

30. Jackson, D. E. *et al.* Superconducting and magnetic phase diagram of $RbEuFe_4As_4$ and $CsEuFe_4As_4$ at high pressure. *Phys. Rev. B* **98**, 014518 (2018).

31. Rojas, D. P., Rodríguez Fernández, J., Espeso, J. I. & Gómez Sal, J. C. Antiferromagnetic Kondo lattice behaviour of $YbNiAl_2$ alloy. *J. Alloys Compd.* **502**, 275–278 (2010).

32. Butschkow, C. *et al.* Origin of negative magnetoresistance of GaAs/Ga, Mn)as core-shell nanowires. *Phys. Rev. B* **87**, 275–278 (2013).

33. Cai, W. *et al.* Pressure-induced superconductivity and structural transition in ferromagnetic $CrSiTe_3$. *Phys. Rev. B* **102**, 144525 (2020).

34. Hirofumi, W. *et al.* Thermal Expansion and Electrical Resistivity of $EuNi_2(Si_{1-x}Ge_x)_2$. *J. Phys. Soc. Jpn*. **68**, 950-953 (1999).

35. Shunsuke, S. *et al.* Spintronic superconductor in a bulk layered material with natural spin-valve structure. arXiv: 2001.07991 (2020).
36. Zhao, L. *et al.* Monoclinic EuSn$_2$As$_2$ : A Novel High-Pressure Network Structure. arXiv: 2102.00437